# A water film motor

R. Shirsavar, A. Amjadi, N. Hamedani Radja, M. D. Niry,
M. Reza Rahimi Tabar[†] and M. R. Ejtehadi*

*Department of Physics, Sharif University of Technology, P.O. Box 11365-9161, Tehran, Iran*
[†]*and CNRS UMR 6202, Observatoire de la Cote d'Azur, BP 4229, 06304 Nice Cedex 4, France*

We report on electrically-induced rotations in water films, which can function at many length scales. The device consists of a two-dimensional cell used for electrolysis of water films, as simple as an insulator frame with two electrodes on the sides (Fig. 1a), to which an external in-plane electric field ***E***, perpendicular to the mean electrolysis current density ***J***, is applied. If either the external field or the electrolysis current (voltage) exceeds some threshold (while the other one is not zero), the liquid film begins to rotate.

If viscous dissipation in the quasi-two-dimensional film (e.g., one in a square frame) balances the power induced by the current ***J*** and the electric field ***E***, one finds a quasi-stationary two-dimensional (2D) vortex. Dimensional analysis suggests that a factor of |***E***|·|***J***| is the applied power density to rotate the film. Since the rotations start when this power exceeds a constant threshold, we expect an inverse linear relation between the threshold values of ***J*** and ***E***. The measured external-field thresholds for starting the rotations at several electrolysis voltages, which are proportional to J, are in agreement with this argument (Fig. 1b). Our experiments also show that the film's mean angular velocity is in the direction of ***E*** × ***J***.

We call the device "water film motor," as the direction and speed of the rotation are controlled through the direction and strength of the current and/or external electric field. The movies demonstrating the experiment are available in http://mehr.sharif.edu/~softmatter/FilmMotor.

Although the chemistry of water electrolysis and the physics of liquid films have both been studied for a long time, we are still far from having a full understanding of both[1]. The experiments reported here, in addition to their potential applications (e.g., experimental investigations of quasi-2D turbulence[2], separation of binary fluid mixtures at meso- and nanoscales, etc.) can also provide us with a new physical probe for deepening our understanding of water electrolysis and the physics of film flows.

Various mechanisms may produce sufficient torque in the system to rotate the film. One is nonuniform ion distribution, caused by the external electric filed, that can break the symmetry under translation of the frictional forces acting on the fluid molecules. This predicts the correct direction of rotation, but does not answer the question of why adding salt to pure water does not change the rotation speed significantly.

The second possible mechanism is based on ionization of the water molecules in the thin layers close to the electrodes. The oppositely-charged ions are attracted by the electrodes while the remaining ions near the two electrodes interact with the external electric field. This generates two anti-parallel water jets close to the electrodes which create a torque on the film, again in the correct

(observed) direction. In this picture, the source of rotation lies in the boundary, while in the stationary state one should also have a vortex with a uniform angular velocity. However, our measurements (Fig. 1c) indicate that the angular velocity is a monotonically- and slowly-decreasing function of the radius. Thus, although the two aforementioned mechanisms can rotate the water film and predict the rotations' direction, the deviations of our experimental data for the angular velocity from what the two mechanisms predict suggest the possible presence of a third, more complex mechanism in the system that can rotate the films.

For instance, reorientation of the water molecules through diffusion of hydrogen bonds leads to molecular-scale rotation[3]. The diffusion process is enhanced in the electrolysis process (due to the external field), while interactions between H2O dipoles and the applied external field may break the isotropy in the angular jumps, hence giving rise to rotations at larger scales.

Further work and confirmation of the experiments at still smaller length scales may lead to the smallest molecular motor, namely, water films at molecular scales.

We would like to thank M. Sahimi for reviewing the manuscript and his valuable comments. We also thank S. Rahvar for helping us in experiments data analyzing.

## Figure caption:

(a) The set up of the experiment. The external electric field is in the film's plane. The motor works perfectly with just pure water, but to increase the film's life time, we dissolve some glycerin and a little detergent in the water. In this way we make long stable films with microscale thicknesses that rotate up to several minutes before they pop up (for movies demonstrating the experiments, see the supplementary materials). Depending on the aspect ratio of the film frame, the applied external electric field may produce several vortices with different radii.

(b) Logarithmic plot of the electric field's threshold for rotation versus electrolysis voltage indicates a slope of -1.00 ± 0.02, which is in agreement with our simple dimensional analysis. The graphs show the results for different volume fractions of the glycerin, $\phi_g$, in the solution, 0.1 (▲), 0.3 (■) and 0.5 (●).

(c) The film's angular velocity decreases monotonically with the radial distance from the vortex center in $\phi_g$ = 0.06 glycerin solution. We have waited for 45 seconds (■) and 7 minutes (▲) after the rotation has started.

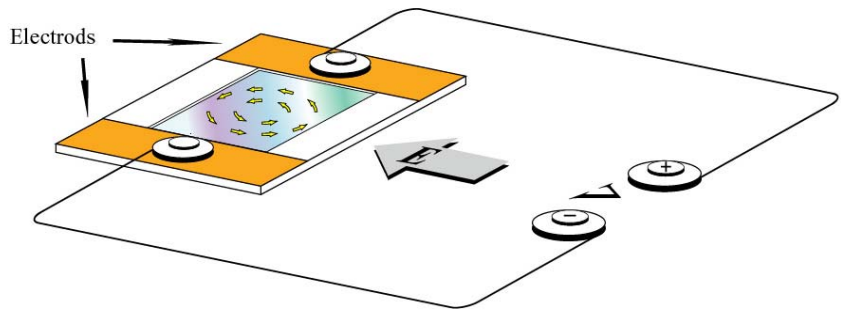

Fig 1a

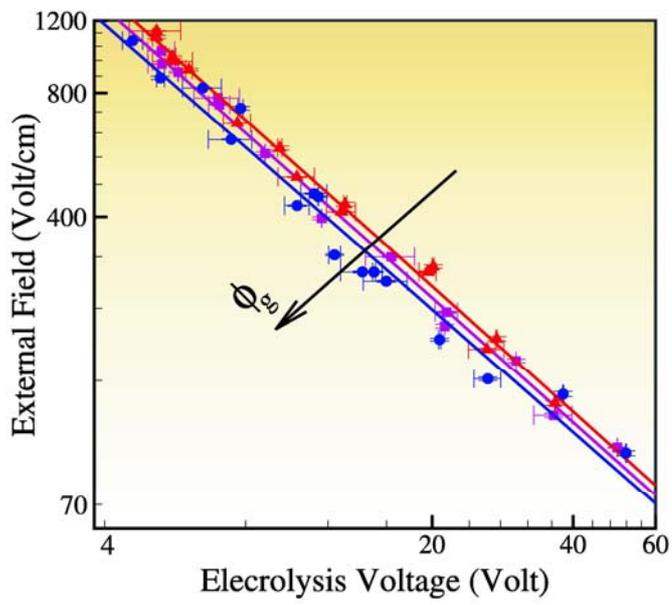

Fig 1b

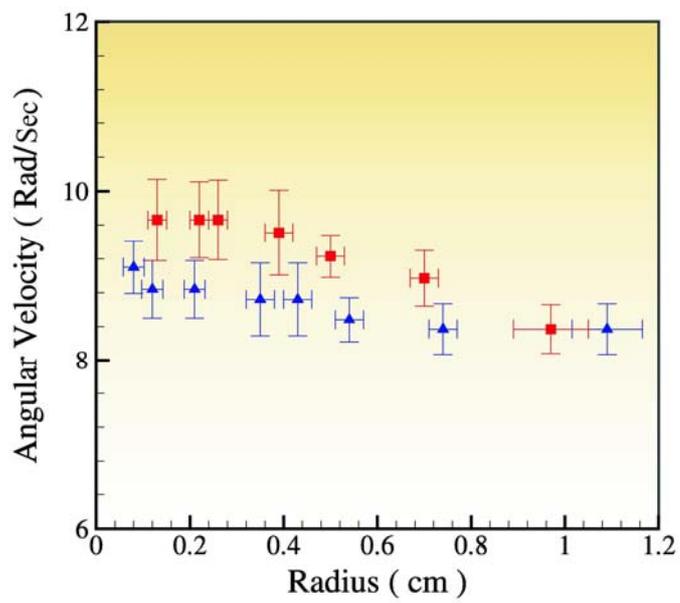

Fig 1c